\documentclass[a4paper,11pt]{article}

\usepackage{latexsym}
\usepackage{amsfonts}
\usepackage{color}
\usepackage{bbm}
\usepackage{amsmath}
\usepackage{amssymb}
\usepackage{tabularx}
\usepackage{chicagoz}
\usepackage{theorem}
\usepackage{pstricks,pst-plot}
\usepackage{graphicx}
\usepackage{epsfig}
\usepackage{lscape}
\usepackage{rotating}
\usepackage{float}
\usepackage{cite}
\usepackage{caption}
\usepackage{tabulary}
\usepackage{hyperref}

\author{Eduardo Ramos-P\'erez$^{(1)}$,\\ Pablo J. Alonso-Gonz\'alez$^{(2)}$, Jos\'e Javier N\'u\~nez-Vel\'azquez$^{(2)}$ }

\title{Mack-Net model: Blending Mack's model with Recurrent Neural Networks}

\begin{document}
\maketitle
\begin{abstract}
\noindent
In general insurance companies, a correct estimation of liabilities plays a key role due to its impact on management and investing decisions. Since the Financial Crisis of 2007-2008 and the strengthening of regulation, the focus is not only on the total reserve but also on its variability, which is an indicator of the risk assumed by the company. Thus, measures that relate profitability with risk are crucial in order to understand the financial position of insurance firms. Taking advantage of the increasing computational power, this paper introduces a stochastic reserving model whose aim is to improve the performance of the traditional Mack's reserving model by applying an ensemble of Recurrent Neural Networks. The results demonstrate that blending traditional reserving models with deep and machine learning techniques leads to a more accurate assessment of general insurance liabilities. 
\end{abstract}
{\small
}

\section{Introduction}
\label{introd}
As an accurate estimation of future payments and its volatility allows the management to take correct underwriting and reinsurance decisions, reserving, understood as the calculation of the amount of required reserves for insurance policies, plays a fundamental role in general insurance firms. The interest of investors and regulators in analysing the volatility of financial institutions has increased significantly since the 2007-2008 Financial Crisis. Regulatory requirements have been enhanced with laws such as Solvency II Directive and Solvency Swiss Test in order to assess the risk profile of insurance companies. Since then, investors are not only focused on the profit but also on the level of risk assumed by the insurance firm to obtain it. Thus, indicators that relate profitability with risk such as the `Return on Risk Adjusted Capital' (\shortciteNP{Braun_2018}) have increased remarkably their influence on the stock prices of financial institutions.\\

Taking into consideration the historical information, the first reserving methods for estimating the ultimate cost in non-life insurance were only focused on obtaining the most likely scenario. Therefore, these deterministic models were unable to estimate the loss reserve uncertainty. Chain Ladder is the most widely used methodology within this family of reserving models. Nevertheless, \citeN{BF_1972} model tends to perform better when historical information is not stable enough to apply the Chain Ladder methodology.\\

As stated previously, general insurance firms are not only interested in the expected ultimate cost but also in its volatility. Consequently, different stochastic methodologies linked to the Chain Ladder procedure were developed. One of the most popular models for estimating the loss reserve variability was introduced by \citeN{Mack_1993}. This methodology, commonly known as Mack's model in the literature, derives the reserve variability by focusing on the first two moments. \citeN{england_verrall_2006} developed a bootstrap method that allows the analyst to obtain a complete reserve distribution by applying the free-distribution model of Mack. \\

Another widely used method to calculate reserve variability is the Overdispersed Poisson (ODP) model, which was developed by \citeN{Renshaw_1998}. This method assumes that incremental payments follow an ODP distribution, where their variance is proportional to their mean. In this model, incremental payments must be positive, but this limitation can be overcome by using the quasi-likelihood approach developed by \citeN{mccullagh_nelder_1989}. As in the case of Mack's model, a complete reserve distribution can be obtained by applying the bootstrap procedure suggested by \citeN{england_verrall_1999} and \citeN{england_2002}. \\

For those cases where data does not follow an ODP distribution, there are other methods based on Chain Ladder such as the log-normal model of \citeN{Kremer_1982}, the gamma procedure of \citeN{Mack_1991} and the negative binomial methodology developed by \citeN{Verrall_2000}. The distribution function of some of the former models can be obtained by using Bayesian inference. \citeN{england_verrall_2006} introduced the procedure for implementing a Bayesian ODP, Mack and Negative Binomial models. The computation of the loss reserve distribution by means of Bayesian inference was recently expanded by \citeN{Meyers_2015}, who introduced several Bayesian Markov Chain Monte-Carlo (MCMC) models for incurred and paid data. These models (Levelled Chain-Ladder, Correlated Chain-Ladder, Levelled Incremental Trend, Correlated Incremental Trend and Changing Settlement Rate) have the aim of improving the performance of the traditional models based on Chain Ladder. This is achieved by including effects such as recognizing the correlation between accident years, applying a skewed distribution to negative incremental payments, introducing a trend over the different development years and implementing the possibility of applying changes in the claim settlement rate.\\

As incurred and paid data can have different patterns and characteristics, there is a set of loss reserving models, based on the Chain Ladder methodology, which were developed in order to take into consideration both data sources. The most relevant methods within this family are Munich Chain Ladder (\citeNP{QM_2004}), Double Chain Ladder (\shortciteNP{MNV_2012}) and Paid-Incurred Chain (\shortciteNP{PCV_2008}). With regard to this last model, it is worth mentioning that \citeN{MW_2010} developed a Bayesian implementation of it, while \shortciteN{HMW_2012} and \citeN{HW_2013} introduced methods strongly related to this approach. Finally, \citeN{Hal_2009} and \citeN{Ven_2008} used incurred and paid data to develop regression-based reserving models, while \shortciteN{PAD_2014}, \citeN{AP_2014} and \shortciteN{MNV_2013} used both sources of information to estimate the expected ultimate cost.\\

The increase of the computational power and the success of machine and deep learning in many fields (\shortciteNP{LCY_2015}, \shortciteNP{SHA_2016}, \shortciteNP{SHS_2017}, \shortciteNP{GPT3_2020} and \shortciteNP{RAN_2021_2}) have facilitated the formation of a new family of reserving models based on these techniques. \citeN{GW_2018} and \citeN{W_2018} applied Artificial Neural Networks (ANN) to predict claim reserves, while \shortciteN{LMT_2019}, \citeN{BR_2019} and \citeN{W2_2018} used a tree-based algorithm, extremely randomized trees (\shortciteNP{GEW_2006}) and regression trees respectively for that purpose. \shortciteN{GRW_2018} and \citeN{GA_2019} demonstrated that it is possible to embed traditional Chain Ladder techniques (ODP model) into a neural network framework. This algorithm was also used by \citeN{Kuo_2018} in order to predict the expected future payments. \shortciteN{RAN_2021} combined ANNs with Random Forests (\citeNP{Breiman_2001}) and Gradient Boosting with regression trees (\citeNP{Friedman_2000}) in order to predict general insurance reserves. In addition to the previous reserving models, \shortciteN{DTM_2011} applied support vector machines to classify data in homogeneous groups of risks before the reserve calculation.\\

The stochastic reserving model presented in this paper (Mack-Net) combines Recurrent Neural Networks (\shortciteNP{RH_1986}) with Mack's model in order to produce more accurate reserve predictions and risk measures. For each individual triangle, an ensemble of Recurrent Neural Networks (RNNs) is fitted in order to forecast both the future payments and the Mack's model parameters. In a second stage, a bootstrap method based on Mack's model is combined with the former predictions in order to compute a full reserve distribution. Consequently, the proposed model has the aim of improving the performance of Mack's model by applying RNNs, which can learn more features than the Chain Ladder technique. Mack-Net model differs from other methods in many ways but the two main differences are explained. First of all, most of the existing reserving models based on machine and deep learning does not produce an estimate of the reserves variability. The few of them that can produce this estimation need to assume a pre-defined theoretical distribution for the payments or incurred cost. Nevertheless, the suggested methodology produces a full reserve distribution without considering any assumption about the payments or incurred cost distribution. Second, information from the same portfolios of several entities tend to be used for fitting reserving models based deep or machine learning techniques. As suggested by regulations like Solvency II Directive, actuaries must aggregate data in homogeneous risk groups, leading to a situation where individual companies do not have available several triangles with similar characteristics to fit the former models. Besides regulatory requirements, if individual companies split triangles with similar characteristics into different pieces, the resulting triangles will not be enough robust in most of the cases. As only one triangle is needed to fit the Mack-Net model, this problem is not present in the suggested methodology. The implementation of the MackNet model in R, the database used for fitting the models and code examples are available in \href{https://github.com/EduardoRamosP/MackNet}{https://github.com/EduardoRamosP/MackNet}\\ 

The proposed model has the aim of producing a more appropriate risk estimation and reserve distribution than other stochastic reserving approaches. Regulations such as Solvency II, Swiss Solvency Test or IFRS promote the use of stochastic models. For example, Risk Adjustment of IFRS 17 has to be based on a certain percentile of the reserve distribution and Reserving Risk of Solvency II Directive can be derived from an stochastic reserving model. Therefore, the calculation of an accurate reserve distribution can lead to lower solvency requirements and less liabilities (Risk Adjustment). It is also worth mentioning that an accurate estimation of the risk profile can lead to a more efficient risk strategy, better portfolio management actions and, therefore, an optimization of the profit-risk indicators. Since the Financial Crisis of 2007-2008, the valuation of financial institutions is not only based on the future profit but also on its volatility or uncertainty. Nowadays, profit-risk indicators are particularly relevant for the valuation of financial institutions.\\

The rest of the paper proceeds as follows: Section \ref{Benchmark_Validation} defines the validations metrics and models used as benchmark to assess the performance of the suggested method. In Section \ref{stack}, the theoretical background and architecture of the Mack-Net model are explained. Empirical results of the benchmark and proposed model are shown in Section \ref{resul}. Finally, Section \ref{conc} presents the main conclusions drawn from the results shown in Section \ref{resul}.\\

\section{Benchmark model and validation metrics}
\label{Benchmark_Validation}

\subsection{Benchmark model}
\label{Benchmark}

This paper presents an extension, based on RNNs, of the traditional Mack's model. Thus, this approach (\citeNP{Mack_1993}) and its bootstrap implementation (\citeNP{england_verrall_2006}) will be used as benchmark for validating the proposed model.\\

The main characteristic of this model compared to others based on Chain Ladder is the lack of assumptions about the underlying distribution of the payments. Mack's model assumes that cumulative payments, \(D_{ij}\), or incurred cost have the following variance and expected value:
\begin{align}
&E[D_{ij}]=\hat{f}_{j} D_{i,j-1}     &Var[D_{ij}]= \hat{\sigma}^2_j D_{i,j-1}
\end{align}
where \(i=(1,2,\dots,I)\) indicates the accident or underwriting year and \(j=(1,2,\dots,I)\) the development year. As explained by \citeN{Mack_1993}, the parameters of the previous expressions are calculated as follows:
\begin{align}
&\hat{f}_{j}=\frac{\sum^{I-j+1}_{i=1}{D_{ij}}}{\sum^{I-j+1}_{i=1}{D_{i,j-1}}}     &\hat{\sigma}^2_{j}=\frac{1}{I-j-1}\sum^{I-j+1}_{i=1}{D_{i,j-1} \bigg( \frac{D_{ij}}{D_{i,j-1}}-\hat{f}_j \bigg)^2}
\end{align}
where \(\{\hat{f}_{j}: j=(2,3,\dots,I)\}\) and \(\{\hat{\sigma}^2_{j}: j=(2,3,\dots,I)\}\). The residuals needed for the bootstrap method (\citeNP{england_verrall_2006}) are calculated as defined below:
\begin{align}
\hat{r}_{ij}=\frac{\sqrt{D_{i,j-1}}*\left(\frac{D_{ij}}{D_{i,j-1}}-\hat{f}_{j}\right)}{\hat{\sigma}_{j}}
\end{align}
To obtain the final residuals, the bias adjustment is added accordingly to the expression suggested by \citeN{england_verrall_2006}:
\begin{align}
\hat{r}_{ij}=\frac{N}{N-p} * \frac{\sqrt{D_{i,j-1}}*\left(\frac{D_{ij}}{D_{i,j-1}}-\hat{f}_{j}\right)}{\hat{\sigma}_{j}}
\end{align}
where \(N\) is the total number of residuals and \(p\) the number of parameters. Hence, the resampled link ratios are obtained as follows:
\begin{align}
f^B_{ij}=\hat{f}_{j}+ r^B_{ij} \frac{\hat{\sigma}_{j}}{\sqrt{D_{i,j-1}}}
\end{align}
where \(B\) refers to the number of upper triangles to be simulated and \(r^B_{ij}\) to the residual resampled in the position (\(i,j\)) of the \(B^{th}\) triangle. Taking into consideration \(D_{i,j}\) and the resampled link ratios, a new set of development factors, \(\tilde{f}^B_{j}\), is computed. Typically, a zero mean adjustment is applied to the residuals in order to ensure that the mean of the stochastic process is the same as the deterministic Chain Ladder method, which is fully dependent on \(\hat{f}_{j}\). \\

The lower triangle (\(D_{i,j}\) where \(i+j > I+1\)) is predicted by combining \(\tilde{f}^B_{j}\) and the upper triangle (\(D_{i,j}\) where \(i+j \leq I+1\)). Then, the process variance is incorporated to the lower triangle by adding the following expression: \(\hat{\sigma}_{j}r^B_{ij}\sqrt{D_{i,j-1}}\). In case further details about the bootstrap method are needed, refer to \citeN{england_verrall_2006} and \citeN{J_2011}.\\

Although the methodology proposed in this paper merges RNNs with Mack's model, other two approaches have been selected as benchmark: Staked-ANN (\shortciteNP{RAN_2021}) and Changing Settlement Rate. The first model combines ANNs with Random Forests and Gradient Boosting with regression trees (\citeNP{Friedman_2000}) to predict general insurance reserves. The stochastic procedure of Stacked-ANN assumes that payments follow a log-normal distribution. On the other hand, Changing Settlement Rate (CSR) is a Bayesian Markov Chain Monte-Carlo model. The prior distributions and the simulation approach are proposed by \citeN{Meyers_2015}.

\subsection{Validation metrics}
\label{metrics}
General insurance companies are asked by insurance regulations like Solvency II Directive and Solvency Swiss Test to evaluate their reserve variability. Thus, the accuracy and variability of Mack-Net model (Section \ref{stack}) will be validated and compared with the benchmark model defined in Section \ref{Benchmark}.\\

To assess the accuracy of the reserve predicted by the models, the following error measure will be computed for every line of business:
\begin{align}
\%RMSE(U^t)=\sqrt{\frac{1}{K}\sum^K_{n=1}\left(\frac{\hat{U}^t_n-U^t_n}{U^t_n}\right)^2}*100
\end{align}
\begin{align}
\%MAE(U^t)=\frac{100}{K}\sum^K_{n=1} \left| \frac{\hat{U}^t_n-U^t_n}{U^t_n} \right|
\end{align}
where \(K\) is the total number of companies analysed, \(\hat{U}^t_n\) the ultimate cost predicted by the reserving model for the \(n^{th}\) company and \(U^t_n\) the ultimate cost that was actually observed. The Model Confidence Test (\shortciteNP{HLN_2011}) will be also applied to produce a more robust comparison of models accuracy. This procedure consists on a sequence of tests which permit the identification of the best models at a certain confidence level.\\

In addition to the previous error measures, the reserve variability produced by the different stochastic reserving models will be assessed by applying the statistical test introduced by \citeN{Kupiec_1995}. The aim of this test is to validate the Value-at-Risk (VaR) by comparing the number of VaR breaches with the percentile selected for calculating the VaR. In this paper, the percentile selected for evaluating the reserve variability is \(\alpha=0.995\), which is the level set up by Solvency II to calculate the risk of insurance companies. The empirical results of the test and \(\%RMSE(R^t)\) are collected in Section \ref{Comparison}.\\

\section{Data and Mack-Net architecture}
\label{stack}
\begin{figure}[!htb]
\begin{center}
\caption{Mack-Net model architecture}
\includegraphics[width=0.99\textwidth]{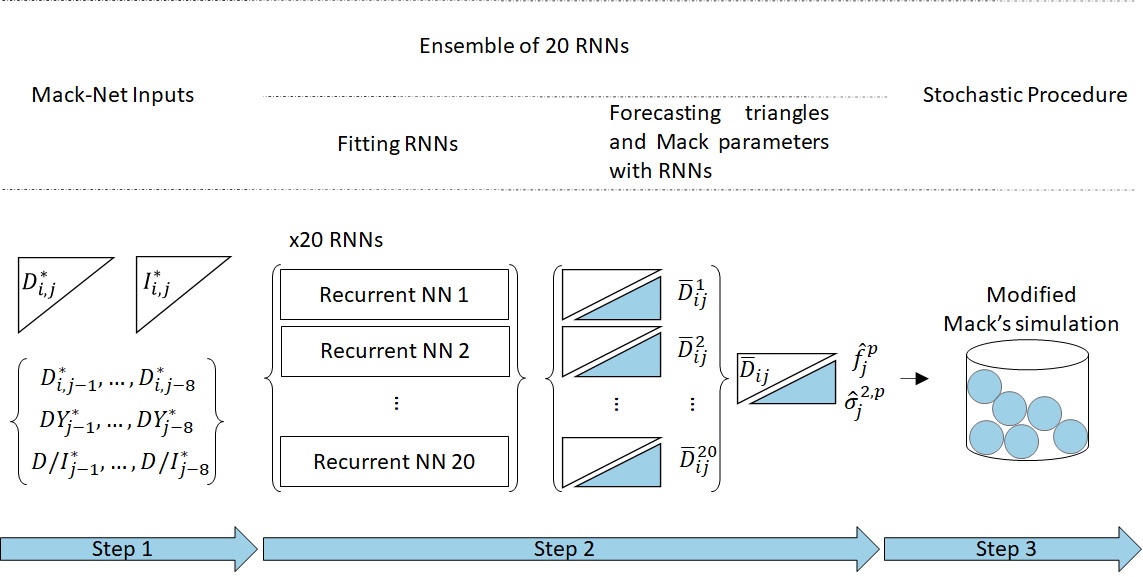}
\label{fig:Fig 1}
\end{center}
\end{figure}

The aim of this section is to explain the architecture of the Mack-Net model. To do so, this section has been divided into three different subsections. The inputs of the model are described in the first one, the ensemble of RNNs is explained in the second subsection and, in the last one, the bootstrap method to obtain a reserve distribution is presented. To support the explanation, Figure \ref{fig:Fig 1} shows the architecture of the proposed stochastic reserving model.\\

\subsection{Data source and model inputs}
\label{data}
The starting point of the Mack-Net model is the definition of inputs used within the ensemble of RNNs. Hence, the goal of this subsection is to define the database used, as well as the response and explicative variables for fitting the RNNs ensemble.\\

The database of Schedule P of the NAIC Annual Statement (available on \href{https://www.casact.org/research/index.cfm?fa=loss_reserves_data}{CAS website}) is selected for fitting and validating the Mack-Net model. The paid data, incurred cost and premiums available in the previous database were collected from general insurance companies from the US. The range of accident years contained in the Schedule P database is 1988-1997. As the upper and lower triangles are included, ten development years are available for each accident year.\\

The benchmark and the proposed model will be fitted to the 200 loss triangles selected by \citeN{Meyers_2015} from the Schedule P database. This author pointed out that one of the main mistakes that could be made with NAIC Schedule P data is selecting triangles from insurance companies that have experienced significant changes in business operations. Therefore, the net-on-gross premiums ratio and the coefficient of variation of the net premiums were used by Meyers to identify those companies that made changes in their business operations or reinsurance structure. Taking into consideration these indicators, Meyers selected 50 loss triangles from each of the following lines of businesses: Commercial Auto (CA), Private Passenger Auto Liability (PA), Workers' Compensation (WC) and Other Liability (OL). The codes of the companies selected can be found in \citeN{Meyers_2015}.\\

It is worth mentioning that loss triangles are considered the primary method to organize the observed incurred cost or payments for general insurance reserving purposes. Loss triangles show the total losses of different underwriting or accident years at various valuation dates. This data shows the profitability and behaviour of the portfolios. Thus, this data is normally not available because insurance entities do not make public this information. Schedule P database is used in the academic field by several authors (such as \shortciteNP{Leong_2014}, \citeNP{Meyers_2015}, \citeNP{Kuo_2018} or \shortciteNP{RAN_2021}) because it offers the possibility of testing triangles from numerous companies and different lines of business.\\

Before starting with the definition of the explanatory and response variables, it is worth mentioning that the last diagonal of the triangle is selected as a test set for fitting the ensemble of RNNs. Thus, the remaining triangle (9 development years) is used to train the algorithms. The sequences used as explanatory variables, \(X_i\), and the response variable, \(Y\), are the following:
\begin{gather}
Y = C^*_{ij} = \frac{C_{ij}}{P_{i}} \\
X_1 = \left( C^*_{ij-1}, C^*_{ij-2}, \dots , C^*_{ij-8} \right) = \left( \frac{C_{ij-1}}{P_{i}}, \frac{C_{ij-2}}{P_{i}}, \dots , \frac{C_{ij-8}}{P_{i}} \right) \\
X_2 = \left( DY^*_{j-1}, DY^*_{j-2}, \dots , DY^*_{j-8} \right) = \left( \frac{DY_{j-1}}{I}, \frac{DY_{j-2}}{I}, \dots , \frac{DY_{j-8}}{I} \right) \\
X_3 = \left( R^*_{j-1}, \dots, R^*_{j-8} \right)= \left( \frac{\sum^{I-j+2}_{i=1}{D^*_{ij-1}}}{\sum^{I-j+2}_{i=1}{\frac{IC_{ij-1}}{P_i}}}, \dots, \frac{\sum^{I-j+9}_{i=1}{D^*_{ij-8}}}{\sum^{I-j+9}_{i=1}{\frac{IC_{ij-8}}{P_i}}} \right) 
\end{gather}

where \(P_{i}\) is the premium, \(C_{ij}\) is the incremental payment, \(D^*_{ij}\) is equal to \(D_{ij} / P_i\), \(I\) the total number of accident years and \(IC_{ij}\) represents the incurred cost of the accident year \(i\) and development \(j\). \(R^*_{j}\) is the scaled cumulative payments between the scaled incurred cost. Thus, \(X_{1}\), \(X_{2}\) and \(X_{3}\) are the time series used as input in the RNNs to predict the next year payment. Two remarks about the variables need to be made:

\begin{enumerate}
\item Payments and development years were scaled. \(DY_j\) were initialized as one and then scaled to the range \([0,1]\). To do so, \(DY_j\) were divided between the total number of accident years, \(I\). In the case of payments, premiums (\(P_{i}\)) play the role of exposure measure. The use of exposure measures is widely used in reserving models, otherwise the changes in the claims settlement speed will be mixed with variations in the business volume.
\item As it can be observed in the formal definition of the explanatory variables, the last \(8\) observations of each variable are selected. This number was chosen due to the size of the triangles. Ten development years are available but only \(9\) of them are used during the training of the RNNs. The other development year is used as a test set. Thus, as the aim of the RNNs is to predict the value in \(t\), the maximum lag available for training the algorithms is \(t-8\). Each RNN forecasts the next payment by taking into consideration the information available in the last 8 periods.
\end{enumerate}

Before concluding this subsection it is worth mentioning that the model can be applied to predict the incurred cost. To do so, \(X_1\) and \(Y\) should be substituted by the following expressions:
\begin{gather}
Y' = i^*_{ij} = \frac{i_{ij}}{P_{i}} \\
X'_1 = \left( IC^*_{ij-1}, IC^*_{ij-1}, \dots , IC^*_{ij-8} \right) = \left( \frac{IC_{ij-1}}{P_{i}}, \frac{IC_{ij-2}}{P_{i}}, \dots , \frac{IC_{ij-8}}{P_{i}} \right)
\end{gather}
where \(i_{ij}\) is the incremental incurred cost. The method and calculations that will be explained in Section \ref{rnn} and \ref{stochastic} are the same regardless of the variable to be predicted (payments or incurred cost).

\begin{figure}[!htb]
\begin{center}
\caption{Recurrent Neural Network Architecture}
\includegraphics[width=0.8\textwidth]{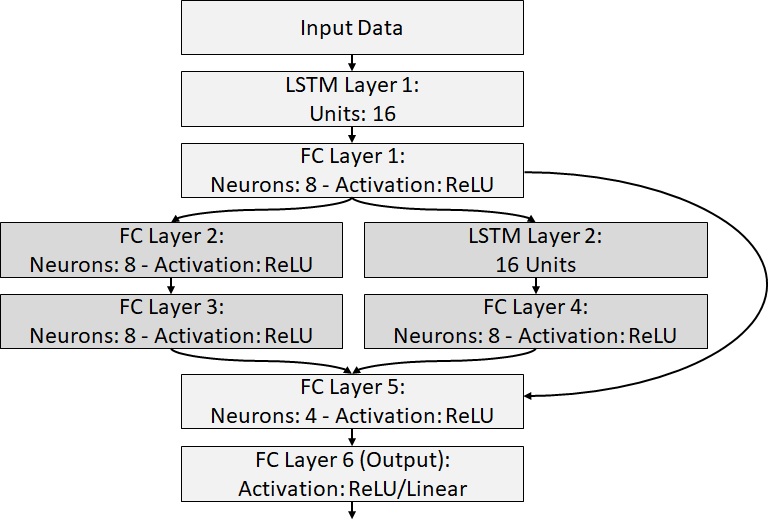}
\label{fig:Fig 2}
\end{center}
\end{figure}

\subsection{Ensemble of RNNs}
\label{rnn}

As shown in Figure \ref{fig:Fig 1}, once the model inputs are prepared, \(20\) Recurrent Neural Networks composed of several Fully Connected (FC) and Long-Short Term Memory (LSTM) layers (Figure \ref{fig:Fig 2}) are fitted. The number of Recurrent Neural Networks fitted is high enough to obtain the average model prediction regardless their initial weights. This strategy was also applied by \citeN{Kuo_2018}.\\

It has to be pointed out that a skip connection between `FC Layer 1' and `FC Layer 5' has been included. This type of connection, introduced by \shortciteN{HZRS_2016} in the field of image recognition with Convolutional Neural Networks, gives the possibility to skip the training of a part of the Neural Network architecture. During the learning process, the RNN will decide by itself if `FC Layer 3' and/or `FC Layer 4' send information to `FC Layer 5'. Apart from giving to the network the possibility to simplify the structure by skipping layers, this kind of connection helps to avoid the problem of vanishing gradients by using the activation of a previous layer until the skipped one learns its weights. \\

\begin{figure}[!htb]
\begin{center}
\caption{LSTM structure}
\includegraphics[width=0.7\textwidth]{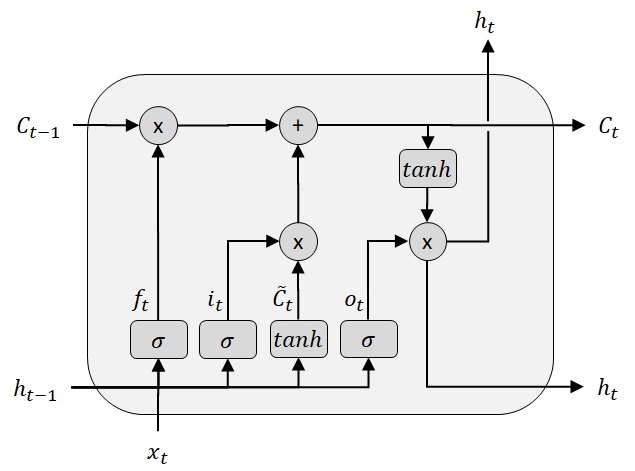}
\label{fig:Fig 3}
\end{center}
\end{figure}

To take temporal dependencies into consideration, the first layer of every RNN is a LSTM cell. This structure was introduced by \citeN{HSJ_1997} for managing time series. Figure \ref{fig:Fig 3} and the following expressions define the LSTM architecture:
\begin{gather} 
f_t=\sigma\left(W_f[h_{t-1},x_{t}]+b_f\right) \\
i_t=\sigma\left(W_i[h_{t-1},x_{t}]+b_i\right) \\
\tilde{C}_t=\tanh\left(W_c[h_{t-1},x_{t}]+b_c\right) \\
C_t=f_t C_{t-1} + i_t \tilde{C}_t\\
o_t=\sigma\left(W_o[h_{t-1},x_{t}]+b_o\right) \\
h_t=o_t \tanh(C_t)
\end{gather}
Where \(W_f\), \(W_i\), \(W_c\), \(W_o\), \(b_f\), \(b_i\), \(b_c\) and \(b_o\) represent the weights and bias of the RNNs and \(\sigma(x)\) the logistic sigmoid function.\\

The main characteristics of the RNNs are defined below:
\begin{itemize}
\item The algorithm used to optimize the weights is Adaptative Moment Estimation (ADAM), which was developed by \citeN{dk_2014}. Considering current and previous gradients, this procedure allows to implement a progressive adaptation of the initial learning rate. These authors suggested the following default values for the ADAM parameters: \(\beta_1=0.9\) and \(\beta_2=0.999\). In this paper, the initial learning rate is set to \(\delta=0.01\) and the default ADAM parameters are used during the training process.
\item The batch size is equal to the number of observations of the training set
\item The backward pass calculations are done taking the mean squared error as loss function.
\item Each individual algorithm within the ensemble is randomly initialized. Glorot (\citeNP{GB_2010}) initializer is used for the LSTM weights responsible of transforming linearly the inputs, while the LSTM weights for the linear transformations within the recurrent states are initialized with the orthogonal approach suggested by \shortciteN{SMG_2013}. 
\item In order to avoid overfitting, the level of dropout regularization \(\theta\) (\shortciteNP{SHK_2014}) is set to \(5\%\).
\end{itemize}
The training of the RNNs has been implemented using the Keras (\citeNP{keras_2015}) and Tensorflow (\shortciteNP{T_2015}). As previously stated, the initial weights of the RNNs are randomly initialized. Thus, the lower triangle predicted by every single algorithm is going to be different. Once the lower triangles are predicted with each RNN, the Mack-Net parameters are computed with the predicted cumulative payments or incurred cost as follows:
\begin{gather}
\hat{f}^p_j=\frac{\sum^{I}_{i=I-j+2}{\bar{D}_{ij}}}{\sum^{I}_{i=I-j+2}{\bar{D}_{i,j-1}}} \\
\hat{\sigma}^{2,p}_j=\frac{1}{I-j-1}\sum^{I}_{i=1}{\bar{D}_{i,j-1}}\left(\frac{\bar{D}_{ij}}{\bar{D}_{i,j-1}}-\bar{f}_{j}\right)^2
\end{gather}
where \(\bar{f}_{j}\) is equal to \(\sum^{I}_{i=1}{\bar{D}_{ij}}/\sum^{I}_{i=1}{\bar{D}_{i,j-1}}\) and \(\bar{D}_{ij}=\sum^{K}_{k=1}\bar{D}^k_{ij}/K\). In addition, \(\bar{D}^k_{ij}\) represents the cumulative payments of the lower triangle (\(i+j > I+1\)) predicted by \(k^{th}\) RNN and the observed values of the upper triangle (\(i+j \leq I+1\)). \(K\) is the number of RNNs included in the ensemble. Similar to the notation used in Section \ref{Benchmark}, \(\{\hat{f}^p_j: j=(2,3,\dots,I)\}\) and \(\{\hat{\sigma}^{2,p}_j: j=(2,3,\dots,I)\}\). \\

As it can be derived from the model definition, rather than using the traditional Mack parameters described in Section \ref{Benchmark}, Mack-Net model parameters are estimated by taking into consideration the predictions made by the ensemble of RNNs. Therefore, the central scenario of the stochastic Mack-Net model is equal to the reserve predicted by the ensemble of RNNs, while the mean of the traditional Mack's model converge to the reserve estimated with the deterministic Chain Ladder method. As any other general insurance reserving methodology such as Mack's model, CSR or Stacked-ANN, the approach suggested by this paper is valid until a new diagonal of the loss triangle is available.

\subsection{Stochastic procedure}
\label{stochastic}
As previously stated, the bootstrap method of Mack-Net is based on the traditional Mack's model. This last methodology has been selected as reference for producing the full reserve distribution due to the two following reasons. First, Mack's model derives the distribution by focusing on the two first moments. In contrast to most of the stochastic reserving models, this approach does not make any assumption about the theoretical distribution follow by incurred cost or cumulative payments. Changes in policyholders' behaviour, reinsurance structures, regulation and number of policy holders can modify significantly the payments or incurred cost collected by loss triangles. Therefore, a free-distribution model is especially appropriate for insurance companies and regulators because the previous portfolio changes happen quite often in the sector. Second, Mack's model is already applied by insurance companies to comply with regulations such as Solvency II Directive, Swiss Solvency Test or IFRS. Thus, the bootstrap method of Mack-Net is familiar and aligned with the procedures already used by the insurance market.\\

As no assumption about the underlying distribution of cumulative payments or incurred cost is taken, the expected value and variance of Mack-Net model is defined as follows:
\begin{align}
&E[D_{ij}]=\hat{f}^{p}_{j} \bar{D}_{i,j-1}     &Var[D_{ij}]= \hat{\sigma}^{2,p}_j \bar{D}_{i,j-1}
\end{align}
The Mack-Net model uses the predictions made by the ensemble of RNNs to determine the mean and variance of the reserve distribution. By doing so, the forecasting power of deep and machine learning techniques are taken into consideration. The calculation of residuals needed for the bootstrap method is based on the expression provided by \citeN{england_verrall_2006} for Mack's model:
\begin{align}
\hat{r}^{p}_{ij}=\frac{\sqrt{\bar{D}_{i,j-1}}*\left(\frac{\bar{D}_{ij}}{\bar{D}_{i,j-1}}-\bar{f}_{j}\right)}{\hat{\sigma}^{p}_j} 
\end{align}
where \(\{\hat{r}^{p}_{ij}: j=(2,\dots,I);\; i=(1,\dots,I)\}\). As in Mack's model, Mack-Net model is bias adjusted in accordance with the procedure suggested by \citeN{england_verrall_2006}:
\begin{align}
\hat{r}^{p}_{ij}=\frac{N}{N-p} \frac{\sqrt{\bar{D}_{i,j-1}}*\left(\frac{\bar{D}_{ij}}{\bar{D}_{i,j-1}}-\bar{f}_{j}\right)}{\hat{\sigma}^{p}_j} 
\end{align}
As with Mack's model, \(p\) is equal to the number of development factors. Once the set of residuals has been calculated, the resampled upper triangles of link ratios are calculated as follows:
\begin{align}
f^{B,p}_{ij}=\hat{f}^{p}_{j}+ r^{B,p}_{ij} \frac{\hat{\sigma}^{p}_j}{\sqrt{\bar{D}_{i,j-1}}}
\end{align}
where \(B\) refers to the number of upper triangles to be simulated and \(r^{B,p}_{ij}\) to the residual resampled in the position (\(i,j\)) of the \(b^{th}\) triangle. Similar to the Mack's model presented in Section \ref{Benchmark}, a zero mean adjustment should be applied to the residuals in order to avoid deviations between the simulated and theoretical mean. The resampled development factors (\(\tilde{f}^{B,p}_{j}\)) are
\begin{align}
\tilde{f}^{B,p}_{j}=\frac{\sum^{I-j+1}_{i=1}{\bar{D}_{i,j-1}f^{B,p}_{ij}}}{\sum^{I-j+1}_{i=1}{\bar{D}_{i,j-1}}}
\end{align}
It is worth mentioning that the previous calculation is carried out with the upper triangle (\(i+j \leq I+1\)). Then, \(\bar{D}_{i,j-1}\) can be substituted by \(D_{i,j-1}\). The resampled development factors, \(\tilde{f}^{B,p}_{j}\), are used to calculate the lower triangle (\(i+j > I+1\)) by applying the Chain Ladder methodology: 
\begin{align}
\hat{D}_{ij} = \bar{D}_{i,j-1}\tilde{f}^{B,p}_{j}
\end{align}
Then, as well as in the case of Mack's model, the process variance is included in the Mack-Net model in order to take into consideration the randomness in future outcomes:
\begin{align}
\hat{D}_{ij}=\hat{D}_{ij}+\hat{\sigma}^{p}_jr^{B,p}_{ij}\sqrt{\bar{D}_{i,j-1}}
\end{align}
Notice that, as the procedure is applied recursively, \(\bar{D}_{i,j-1}\) is used in the previous equation only when the simulation refers to year after the last diagonal observed. In the rest of the cases, the recurrence is applied and, thus, \(\bar{D}_{i,j-1}\) substituted by \(\hat{D}_{ij-1}\). As explained in Section \ref{Benchmark} for Mack's model, this procedure (\citeNP{england_verrall_2006}) allows the recognition of the process variance, which is the variability in the forecasts of future payments.

\section{Model fitting and results}
\label{resul}
In this section, Mack-Net parameters by line of business and the comparison between the performance of the benchmark and the Mack-Net model are presented.

\subsection{Fitting of Mack-Net model}
\label{metrics2}
This subsection presents Mack-Net parameters by line of business. As stated before, the model has been fitted individually to each of the \(200\) triangles selected by \cite{Meyers_2015} from the Schedule P of the NAIC Annual Statement. For further details about the database refer to Section \ref{data}.\\

As explained in Section \ref{rnn}, once the ensemble of RNNs has been fitted, the Mack-Net architecture proceeds with the calculation of \(\hat{f}^{p}_{j}\) and \(\hat{\sigma}^{2,p}_j\). Similar to \(\hat{f}_{j}\) and \(\hat{\sigma}^{2}_j\) in Mack's model, Mack-Net parameters define the variance and expected value of the cumulative payments or incurred cost. Thus, Tables \ref{DevFactors}, \ref{DevFactorsII}, \ref{Sigma} and \ref{SigmaII} present a comparison between the average Mack and Mack-Net parameters by line of business.\\

\begin{table}[H]
  \begin{center}
    \caption{Average \(\hat{f}^{p}_{j}\) and \(\hat{f}_{j}\) by line of business (Paid data).}
    \label{DevFactors}
    \begin{tabular}{c c c c c c c c c}
      \hline
                 &\multicolumn{2}{c}{CA}&\multicolumn{2}{c}{PA}&\multicolumn{2}{c}{WC}&\multicolumn{2}{c}{OL}\\
      \hline
      Dev        &      & Mack &      & Mack &      & Mack &      & Mack \\
      Year       & Mack & Net  & Mack & Net  & Mack & Net  & Mack & Net  \\
      \hline
      1          & 1.90 & 1.88 & 1.77 & 1.75 & 2.21 & 2.59 & 3.12 & 3.10     \\ 
      2          & 1.35 & 1.35 & 1.22 & 1.22 & 1.29 & 1.38 & 1.72 & 1.67     \\ 
      3          & 1.16 & 1.15 & 1.10 & 1.10 & 1.13 & 1.15 & 1.34 & 1.30     \\ 
      4          & 1.08 & 1.06 & 1.06 & 1.05 & 1.07 & 1.08 & 1.18 & 1.16     \\ 
      5          & 1.04 & 1.03 & 1.03 & 1.02 & 1.04 & 1.04 & 1.11 & 1.07     \\ 
      6          & 1.02 & 1.01 & 1.01 & 1.01 & 1.02 & 1.02 & 1.04 & 1.03     \\ 
      7          & 1.00 & 1.01 & 1.01 & 1.00 & 1.02 & 1.02 & 1.02 & 1.01     \\ 
      8          & 1.01 & 1.00 & 1.00 & 1.00 & 1.01 & 1.01 & 1.02 & 1.01     \\ 
      9          & 1.00 & 1.00 & 1.00 & 1.00 & 1.01 & 1.01 & 1.01 & 1.00     \\ 
      \hline
     \multicolumn{4}{l}{\emph{Source}: own elaboration}    
    \end{tabular}
  \end{center}
\end{table}

\begin{table}[H]
  \begin{center}
    \caption{Average \(\hat{f}^{p}_{j}\) and \(\hat{f}_{j}\) by line of business (Incurred data).}
    \label{DevFactorsII}
    \begin{tabular}{c c c c c c c c c}
      \hline
                 &\multicolumn{2}{c}{CA}&\multicolumn{2}{c}{PA}&\multicolumn{2}{c}{WC}&\multicolumn{2}{c}{OL}\\
      \hline
      Dev        &      & Mack &      & Mack &      & Mack &      & Mack \\
      Year       & Mack & Net  & Mack & Net  & Mack & Net  & Mack & Net  \\
      \hline
      1          & 1.27 & 1.33 & 1.14 & 1.15 & 1.31 & 1.40 & 1.45 & 1.56    \\ 
      2          & 1.09 & 1.09 & 1.03 & 1.03 & 1.07 & 1.09 & 1.21 & 1.20    \\ 
      3          & 1.03 & 1.03 & 1.01 & 1.01 & 1.02 & 1.03 & 1.07 & 1.07    \\ 
      4          & 1.01 & 1.01 & 1.00 & 1.00 & 1.01 & 1.01 & 1.03 & 1.03    \\ 
      5          & 1.00 & 1.01 & 1.00 & 1.00 & 1.00 & 1.01 & 1.04 & 1.01    \\ 
      6          & 0.99 & 1.01 & 1.00 & 1.00 & 1.00 & 1.01 & 1.01 & 1.02    \\ 
      7          & 1.00 & 1.01 & 1.00 & 1.00 & 1.00 & 1.01 & 1.00 & 1.01    \\ 
      8          & 1.00 & 1.01 & 1.00 & 1.00 & 1.00 & 1.01 & 1.01 & 1.01    \\ 
      9          & 1.00 & 1.01 & 1.00 & 1.00 & 1.00 & 1.01 & 1.00 & 1.01    \\ 
      \hline
     \multicolumn{4}{l}{\emph{Source}: own elaboration}    
    \end{tabular}
  \end{center}
\end{table}
The averages of the development factors of the payment models (Table \ref{DevFactors}) present non-significant differences in most of the cases. The only remarkable differences are the second development factor of OL, and the first and second parameters of WC. Mack-Net development factors are lower than Mack's parameters in all the lines of business with the only exception of Workers' Compensation. This can be proved by computing the multiplicative development factors (product of development factors by line of business and model).\\

With regard to the incurred cost development factors (Table \ref{DevFactorsII}), the differences between both models are minor in the case of CA and PA. However, they become more relevant in the case of WC and OL, especially in the case of the first development year. It is also worth mentioning that Mack-Net development factors are higher than Mack parameters regardless of the line of business.\\

\begin{table}[H]
  \begin{center}
    \caption{Average \(\hat{\sigma}^{2,p}_j\) and \(\hat{\sigma}^{2}_j\) by line of business (Paid data).}
    \label{Sigma}
    \begin{tabular}{c c c c c c c c c}
      \hline
                 &\multicolumn{2}{c}{CA}&\multicolumn{2}{c}{PA}&\multicolumn{2}{c}{WC}&\multicolumn{2}{c}{OL}\\
      \hline
      Dev        &      & Mack &      & Mack &      & Mack &      & Mack \\
      Year       & Mack & Net  & Mack & Net  & Mack & Net  & Mack & Net  \\
      \hline
      1          &13.25 &12.43 &12.96 &11.96 &14.98 &13.94 &26.94 &25.47     \\ 
      2          & 6.82 & 5.96 & 5.73 & 5.09 & 6.06 & 5.43 &10.35 &10.21     \\ 
      3          & 4.43 & 3.76 & 3.61 & 3.29 & 4.07 & 3.70 & 6.67 & 5.58     \\ 
      4          & 2.97 & 2.31 & 2.59 & 2.55 & 3.07 & 2.78 & 4.96 & 4.29     \\ 
      5          & 1.76 & 1.39 & 1.40 & 1.47 & 1.55 & 1.76 & 3.21 & 2.30     \\ 
      6          & 0.98 & 0.74 & 0.87 & 0.79 & 1.46 & 1.19 & 1.95 & 1.23     \\ 
      7          & 0.65 & 0.44 & 0.68 & 0.50 & 1.10 & 0.86 & 0.97 & 0.54     \\ 
      8          & 0.42 & 0.26 & 0.19 & 0.24 & 0.71 & 0.55 & 0.78 & 0.42     \\ 
      9          & 0.32 & 0.17 & 0.15 & 0.15 & 0.53 & 0.36 & 0.44 & 0.18     \\ 
      \hline
     \multicolumn{4}{l}{\emph{Source}: own elaboration}    
    \end{tabular}
  \end{center}
\end{table}

\begin{table}[H]
  \begin{center}
    \caption{Average \(\hat{\sigma}^{2,p}_j\) and \(\hat{\sigma}^{2}_j\) by line of business (Incurred data).}
    \label{SigmaII}
    \begin{tabular}{c c c c c c c c c}
      \hline
                 &\multicolumn{2}{c}{CA}&\multicolumn{2}{c}{PA}&\multicolumn{2}{c}{WC}&\multicolumn{2}{c}{OL}\\
      \hline
      Dev        &      & Mack &      & Mack &      & Mack &      & Mack \\
      Year       & Mack & Net  & Mack & Net  & Mack & Net  & Mack & Net  \\
      \hline
      1          & 7.95 & 7.29 &10.11 & 9.23 &15.13 &13.85 &13.69 &12.67     \\ 
      2          & 4.67 & 4.11 & 4.79 & 4.35 & 9.48 & 8.23 & 9.00 & 8.56     \\ 
      3          & 3.19 & 2.71 & 3.06 & 3.04 & 4.09 & 3.50 & 5.24 & 4.40     \\ 
      4          & 2.29 & 1.96 & 1.62 & 1.57 & 2.69 & 2.19 & 4.01 & 3.58     \\ 
      5          & 1.54 & 1.27 & 1.18 & 1.10 & 2.12 & 1.64 & 3.66 & 2.53     \\ 
      6          & 1.13 & 0.97 & 0.78 & 0.71 & 1.79 & 1.34 & 1.50 & 1.18     \\ 
      7          & 0.77 & 0.68 & 0.31 & 0.38 & 1.30 & 0.94 & 1.15 & 0.78     \\ 
      8          & 0.41 & 0.46 & 0.18 & 0.26 & 0.80 & 0.64 & 0.52 & 0.48     \\ 
      9          & 0.33 & 0.38 & 0.15 & 0.18 & 0.49 & 0.41 & 0.43 & 0.34     \\ 
      \hline
     \multicolumn{4}{l}{\emph{Source}: own elaboration}    
    \end{tabular}
  \end{center}
\end{table}
Before analysing the differences between \(\hat{\sigma}^{2,p}_j\) and \(\hat{\sigma}^{2}_j\), it is worth mentioning that previous cumulative payments or incurred cost play a key role in the model variance, defined in equations \(2\) and \(20\) for Mack and Mack-Net model respectively. Thus, \(\hat{\sigma}^{2,p}_j\) and \(\hat{\sigma}^{2}_j\) have to be analysed by taking into consideration the analysis of the development factors explained in the previous paragraphs.\\

With regard to the models for paid loss data, \(\hat{\sigma}^{2,p}_j\) and \(\hat{\sigma}^{2}_j\) show non-material differences. Nevertheless, it has to be pointed out that Mack parameters are higher than those of Mack-Net in every line of business. As the paid development factors of the Mack model are also higher, the pattern shown in Table \ref{Sigma} reveals that Mack model generates a higher volatility than the proposed methodology.\\

Table \ref{SigmaII} shows that Mack's parameters are higher than those of the Mack-Net model fitted with incurred cost data. In contrast to the models for paid loss data, this effect is partially offset by the Mack-Net development factors that, as shown in Table \ref{DevFactorsII}, are higher than those of the Mack's model.\\

As incurred cost includes the payments and the reserve set up by claim adjusters, this variable should be closer to the ultimate claim cost than the payments. Thus, development factors and reserve volatility should be lower in the case of the models fitted with incurred cost data. The comparison of the average parameters of the models for incurred (Table \ref{DevFactorsII} and \ref{SigmaII}) and paid loss data (Table \ref{DevFactors} and \ref{Sigma}) reveals that this trend is followed by both models.\\

\subsection{Comparison against benchmark models}
\label{Comparison}
This subsection compares the performance of the Mack-Net model with the original methodology proposed by Mack. The variability and accuracy will be compared with the metrics and tests shown in Section \ref{metrics}.\\

As previously explained, the aim of the Mack-Net model is to improve the accuracy of the traditional Mack's methodology by using machine and deep learning algorithms and techniques such as RNNs. Table \ref{Accuracy} and \ref{AccuracyII} show the empirical results of the metrics selected for comparing the models accuracy.

\begin{table}[H]
  \begin{center}
    \caption{\(\%RMSE(U^t)\) by model and line of business}
    \label{Accuracy}
    \begin{tabular}{c c c c c c c}
      \hline
      Line of           & Mack     & Mack-Net & Mack     & Mack-Net & CSR     & Stacked   \\
      business          & Paid     & Paid     & Incurred & Incurred &         & ANN       \\
      \hline
      CA                & 7.98\%   & 6.80\%   & 8.18\%   & 8.03\%   & 9.29\%  & 8.92\%     \\  
      PA                & 6.06\%   & 5.01\%   & 2.62\%   & 4.26\%   & 5.46\%  & 7.78\%     \\ 
      WC                & 7.86\%   & 6.77\%   & 8.15\%   & 6.99\%   & 13.29\% & 7.36\%     \\  
      OL                & 20.20\%  & 17.31\%  & 17.38\%  & 13.48\%  & 27.78\% & 19.80\%    \\ 
      \hline
     \multicolumn{2}{l}{\emph{Source}: own elaboration}    
    \end{tabular}
  \end{center}
\end{table}

\begin{table}[H]
  \begin{center}
    \caption{\(\%MAE(U^t)\) by model and line of business}
    \label{AccuracyII}
    \begin{tabular}{c c c c c c c}
      \hline
      Line of           & Mack     & Mack-Net & Mack     & Mack-Net & CSR     & Stacked   \\
      business          & Paid     & Paid     & Incurred & Incurred &         & ANN       \\
      \hline
      CA                & 5.96\%   & 4.78\%   & 5.46\%   & 5.40\%   & 6.35\%  & 6.86\%     \\  
      PA                & 3.81\%   & 3.44\%   & 1.90\%   & 2.86\%   & 3.68\%  & 3.76\%     \\ 
      WC                & 5.32\%   & 4.60\%   & 5.27\%   & 4.51\%   & 6.01\%  & 4.65\%     \\  
      OL                & 13.41\%  & 12.16\%  & 11.34\%  & 9.88\%   & 18.29\% & 13.47\%    \\ 
      \hline
     \multicolumn{2}{l}{\emph{Source}: own elaboration}    
    \end{tabular}
  \end{center}
\end{table}

With regard to the models for paid loss data, Mack-Net methodology improves the accuracy of the Mack's model in every line of business. \(\%RMSE(U^t)\) decreases by \(14\%\) in WC and OL, \(15\%\) in CA, and \(17\%\) in the case of PA. Similar improvements are also observed in terms of \(\%MAE(U^t)\).\\

The comparison of the \(\%RMSE(U^t)\) and \(\%MAE(U^t)\) obtained from the models for incurred loss data shows that Mack-Net model outperforms the Mack's procedure in all the lines of business with the only exception of PA. It is worth mentioning that the accuracy of the Mack-Net model is especially higher in OL, which is the line of business with the longer duration of liabilities. Thus, an appropriate estimation of reserves is particularly relevant in this case. Empirical results demonstrate that Mack-Net model also outperforms general insurance approached based on Markov Chain Monte-Carlo or machine learning such as CSR or Stacked-ANN.\\

Tables \ref{MSC_RMSE} and \ref{MSC_MAE} show the ranking proposed by Model Confidence Set (MCS) considering \(\%RMSE(U^t)\) and \(\%MAE(U^t)\) as loss functions. This approach confirms the outcomes presented in the previous paragraphs. In fact, Mack-Net Incurred and Paid are ranked as the best and second best model respectively when all the lines of business are considered together (row `Total').\\ 

\begin{table}[H]
  \begin{center}
    \caption{Ranking of models according to MSC (\(\%RMSE(U^t)\) and \(\alpha=0.05\)).}
    \label{MSC_RMSE}
    \begin{tabular}{c c c c c c c}
      \hline
      Line of           & Mack     & Mack-Net & Mack     & Mack-Net & CSR      & Stacked   \\
      business          & Paid     & Paid     & Incurred & Incurred &          & ANN       \\
      \hline
      CA                &\(2^{nd}\)&\(1^{st}\)&\(4^{th}\)&\(3^{rd}\)&\(5^{th}\)&\(6^{th}\)\\  
      PA                &\(6^{th}\)&\(3^{rd}\)&\(1^{st}\)&\(2^{nd}\)&\(4^{th}\)&\(5^{th}\)\\ 
      WC                &\(6^{th}\)&\(1^{st}\)&\(5^{th}\)&\(2^{nd}\)&\(4^{th}\)&\(3^{rd}\)\\  
      OL                &\(5^{th}\)&\(2^{nd}\)&\(3^{rd}\)&\(1^{st}\)&\(6^{th}\)&\(4^{th}\)\\ 
      Total             &\(5^{th}\)&\(2^{nd}\)&\(3^{rd}\)&\(1^{st}\)&\(6^{th}\)&\(4^{th}\)\\ 
      \hline
     \multicolumn{2}{l}{\emph{Source}: own elaboration}    
    \end{tabular}
  \end{center}
\end{table}

\begin{table}[H]
  \begin{center}
    \caption{Ranking of models according to MSC (\(\%MAE(U^t)\) and \(\alpha=0.05\)).}
    \label{MSC_MAE}
    \begin{tabular}{c c c c c c c}
      \hline
      Line of           & Mack     & Mack-Net & Mack     & Mack-Net & CSR     & Stacked   \\
      business          & Paid     & Paid     & Incurred & Incurred &         & ANN       \\
      \hline
      CA                &\(5^{th}\)&\(1^{st}\)&\(3^{rd}\)&\(2^{nd}\)&\(4^{th}\)&\(6^{th}\)\\  
      PA                &\(6^{th}\)&\(3^{rd}\)&\(1^{st}\)&\(2^{nd}\)&\(5^{th}\)&\(4^{th}\)\\ 
      WC                &\(6^{th}\)&\(3^{rd}\)&\(5^{th}\)&\(1^{st}\)&\(4^{th}\)&\(2^{nd}\)\\  
      OL                &\(5^{th}\)&\(3^{rd}\)&\(2^{rd}\)&\(1^{st}\)&\(6^{th}\)&\(4^{th}\)\\ 
      Total             &\(5^{th}\)&\(2^{nd}\)&\(3^{rd}\)&\(1^{st}\)&\(6^{th}\)&\(4^{th}\)\\ 
      \hline
     \multicolumn{2}{l}{\emph{Source}: own elaboration}    
    \end{tabular}
  \end{center}
\end{table}

With regard to the validation of the reserves variability, \citeN{Kupiec_1995} test is applied to assess the appropriateness of the reserve distribution generated by the stochastic process. Table \ref{Risk} collects the p-values of the Kupiec test assuming a VaR percentile of \(\alpha=0.995\), which is the value for evaluating the risk profile of insurance companies under Solvency II Directive.\\

Companies included in each line of business have different volumes. This fact was taken into consideration within the Kupiec test by giving different weights to each company. The higher the standard deviation generated by the company, the higher the weight given to compute the Kupiec test. Table \ref{Risk} collects the p-values by model and line of business.\\

\begin{table}[H]
  \begin{center}
    \caption{Kupiec test (p-values) by model and line of business}
    \label{Risk}
    \begin{tabular}{c c c c c c c}
      \hline
      Line of           & Mack        & Mack-Net    & Mack        & Mack-Net    \\
      business          & Paid        & Paid        & Incurred    & Incurred    \\
      \hline
      CA                & $\geq 0.05$ & $\geq 0.05$ & $<0.05$     & $\geq 0.05$ \\ 
      PA                & $\geq 0.05$ & $\geq 0.05$ & $\geq 0.05$ & $\geq 0.05$ \\ 
      WC                & $<0.05$     & $<0.05$     & $<0.05$     & $\geq 0.05$ \\
      OL                & $\geq 0.05$ & $\geq 0.05$ & $<0.05$     & $\geq 0.05$ \\ 
      \hline
     \multicolumn{2}{l}{\emph{Source}: own elaboration}    
    \end{tabular}
  \end{center}
\end{table}
According to the results of the models for paid loss data, Mack and Mack-Net methodologies are unable to produce an appropriate Value at Risk (VaR) for Workers Compensation. In the rest of lines of business, the excesses of the VaR estimated by both models are aligned with the confidence level selected (\(\alpha=0.995\)). It is worth mentioning that, as discussed in Section \ref{metrics2}, Mack-Net parameters reveal a lower level of variance than those of the Mack's model. Thus, the higher accuracy of the Mack-Net model for paid loss data (Table \ref{Accuracy}) allows to generate appropriate risk measures with a lower level of variability.\\

In the case of the models for incurred cost, Mack-Net model passes the test in all the lines of business, while Mack's model fails the test in three out of four lines of business. As it will be presented in Table \ref{CoV_Percentage}, the coefficients of variation generated by the Mack-Net model are lower than those of the traditional Mack's methodology. Nevertheless, Mack-Net model passes the test because the accuracy of the mean of the stochastic process is higher (Table \ref{Accuracy}).\\

\begin{figure}[!htb]
\begin{center}
\caption{Company code 620. Other Liability.}
\includegraphics[width=1\textwidth]{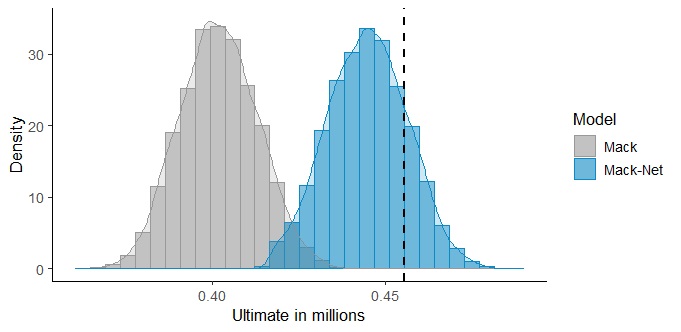}
\label{fig:Fig 4}
\end{center}
\end{figure}

Mack's model fails the test in most of the lines of business due to two reasons. First, the lower level of accuracy leads to higher differences between the actual reserve and the distribution function generated by the model. Second, the model is unable to generate a higher variance in order to offset the lack of accuracy. To illustrate this, Figure \ref{fig:Fig 4} shows one company of the OL segment where Mack's model produces an inappropriate VaR. The observed ultimate is represented by a black dashed line.\\

With the goal of comparing the volatility generated by the different stochastic reserving models, Table \ref{CoV_Percentage} collects the \(\%\) of companies where the coefficient of variation, \(CV(U^t)\), of Mack-Net is lower than in the case of Mack's model:
\begin{table}[H]
  \begin{center}
    \caption{\% of companies where Mack-Net \(CV(U^t)\) \(<\) Mack \(CV(U^t)\)}
    \label{CoV_Percentage}
    \begin{tabular}{c c c c c}
      \hline
      Line of business  & Paid      & Incurred  \\
      business          & loss data & loss data \\
      \hline
      CA                & 56\%      & 72\%      \\  
      PA                & 46\%      & 66\%      \\ 
      WC                & 56\%      & 54\%      \\  
      OL                & 58\%      & 68\%    \\ 
      \hline
     \multicolumn{2}{l}{\emph{Source}: own elaboration}    
    \end{tabular}
  \end{center}
\end{table}
The results shown in tables \ref{Risk} and \ref{CoV_Percentage} demonstrate that Mack-Net model does not need to produce higher coefficients of variation in order to generate more appropriate risk measures than Mack's model. Thus, the proposed methodology produces more efficient risk measures thanks to its predictive power.\\

Finally, on the trade-off between time and accuracy, Table \ref{Time} presents how training time and error change when the input size increases. The database used for fitting the algorithms (Schedule P of the NAIC Annual Statement) contains a maximum range of \(10\) years per each general insurance company. This table sets 5 years of data as the initial scenario and, then, the input size are increased until the maximum available data is reached. The results show that the error decreases in a higher extent than training time in relative terms. This is true even when the number of years is closed to maximum. It is worth noticing that benchmark models accuracy will also decrease if the input size is reduced. Therefore, Table \ref{Time} shows the trade-off between time and accuracy of Mack-Net model in a standalone basis.\\

\begin{table}[H]
  \begin{center}
    \caption{Mack-Net: Training time and error versus input size}
    \label{Time}
    \begin{tabular}{c c c c c c c}
      \hline
                        & 5 Years& 6 Years& 7 Years& 8 Years& 9 Years& 10 Years\\
      \hline
      Time Index        & 100.00 & 102.14 & 105.48 & 113.86 & 126.00 & 140.55  \\  
      Error Index       & 100.00 &  70.73 &  50.05 &  33.73 &  19.02 &   7.08 \\ 
      \hline
     \multicolumn{2}{l}{\emph{Source}: own elaboration}    
    \end{tabular}
  \end{center}
\end{table}

\section{Conclusions}
\label{conc}
The Mack-Net model introduced in this paper has the aim of blending the traditional Mack's reserving model with deep and machine learning techniques. To do so, an ensemble of RNNs is fitted to the loss triangle. Then, the predictions of this ensemble are used for calculating Mack's model parameters. In this paper, the predictive power and reserve variability of the proposed architecture and the traditional Mack's methodology are compared. Models were fitted to 200 incurred cost and paid loss triangles from NAIC Schedule P database (available on \href{https://www.casact.org/research/index.cfm?fa=loss_reserves_data}{CAS website}) in order to generate a robust comparison.\\

Three main conclusions are drawn from the results presented in Section \ref{resul}. First, the comparison of the accuracy reveals that adding deep learning techniques to Mack's model improves the predictive power. With regard to the models fitted with paid data, this paper demonstrates that Mack-Net model outperforms Mack's model in every line of business. In the case of the models for incurred cost, Mack-Net is also more accurate than Mack's model in all the lines of business with the only exception of Personal Auto (PA). The accuracy of reserving models is particularly relevant for long-tail lines of business such as Other Liability (OL). In the case of this last portfolio, Mack-Net methodology reduces the RMSE by \(14\%\) and \(22\%\) when using paid and incurred cost data respectively. Empirical results demonstrate that Mack-Net model also outperforms other reserving approaches based on Markov Chain Monte-Carlo or machine learning such as Changing Settlement Rate or Stacked-ANN respectively.\\

Second, Kupiec test demonstrated that Mack-Net model generates more appropriate risk measures (Value at Risk) than the traditional Mack's methodology. With regard to the paid data, both models fail the test for Workers Compensation (WC). However, in the case of the incurred data, Mack-Net model passes the test in every line of business, while Mack's model fails the test in three out of four lines. Thus, Mack-Net model is not only more accurate but also generates a more appropriate Value at Risk (VaR). The confidence level selected for the VaR is \(\alpha=0.995\), which is the level required by Solvency II to evaluate the reserving risk in insurance companies.\\

Third, the blending of traditional approaches with deep learning techniques generates more efficient models for evaluating the reserving risk, which is the potential cost of deviations from the expected reserve. In other risks such as changes in equity price, the mean of the distribution does not play a relevant role (the mean of the returns are almost always close to zero). As the expected reserve can not be easily predicted, this is not the case for reserving risk, where the appropriateness of the risk measures strongly depends on both the mean and the variance of the reserving model. \\

Due to the reasons explained in the previous paragraph, Mack-Net model is able to generate more appropriate risk measures with a lower variance. Thus, empirical results suggest that the proposed method is more efficient (in terms of risk assessment) than Mack's model because it generates a more reliable VaR with a lower variability.\\

Taking into consideration the previous conclusions, the blending of deep learning techniques with reserving models can be extended to improve the accuracy and risk measures derived from the use of other bootstrapping and Bayesian approaches. In the specific case of the Bayesian reserving models, deep learning algorithms could be applied in order to estimate the parameters of the distributions.\\

\bibliography{Final1}
\bibliographystyle{chicago}
\end{document}